\documentclass[pdflatex,sn-mathphys-num]{sn-jnl}%
\setcitestyle{numbers, square}%
\usepackage{hyperref}%
\usepackage{amsmath,amssymb,amsfonts}%
\usepackage{amsthm}%
\usepackage{mathrsfs}%
\usepackage[title]{appendix}%
\usepackage{xcolor}%
\usepackage{textcomp}%
\usepackage{manyfoot}%
\usepackage{booktabs}%
\usepackage{algorithm}%
\usepackage{algorithmicx}%
\usepackage{algpseudocode}%
\usepackage{listings}%
\usepackage[utf8]{inputenc}%
\theoremstyle{thmstyleone}%
\theoremstyle{thmstyletwo}%
\theoremstyle{thmstylethree}%
\usepackage[all]{nowidow}%

\begin{document}

\title[Article Title]{Quantum Reconstruction and Phenomenology per the Relativity Principle}

\author*[1]{\fnm{W.M.} \sur{Stuckey}}\email{stuckeym@etown.edu}

\author[2]{\fnm{Timothy} \sur{McDevitt}}\email{mcdevittt@etown.edu}
\equalcont{These authors contributed equally to this work.}

\affil*[1]{\orgdiv{Department of Physics}, \orgname{Elizabethtown College}, \orgaddress{\city{Elizabethtown}, \state{PA}, \country{USA}}}

\affil[2]{\orgdiv{Department of Mathematical Science}, \orgname{Elizabethtown College}, \orgaddress{\city{Elizabethtown}, \state{PA}, \country{USA}}}

\abstract{We use the relativity principle to complete axiomatic reconstructions of quantum mechanics (QM) via information-theoretic principles that are based on Darrigol's \textit{discreteness requirement} or its equivalent, e.g., Brukner \& Zeilinger's \textit{Information Invariance \& Continuity} or Khrennikov's \textit{quantum action invariance principle}. In this approach to the quantum reconstruction program (QRP), the Hilbert space kinematics of QM is derived most fundamentally from the experimentally motivated postulate of ``discreteness,'' rendering QM a \textit{principle theory} as defined by Einstein. Special relativity is also a principle theory, since its Lorentz transformation kinematics is derived from the experimentally motivated light postulate (observer-independence of the speed of light $c$). While special relativity has a compelling fundamental principle (relativity principle) to account for its experimentally motivated light postulate, QRP has not produced a compelling fundamental principle to account for its experimentally motivated ``discreteness'' requirement. We complete this particular approach to QRP by showing how the ``discreteness'' requirement is necessitated by the relativity principle and Planck's radiation law, just like the light postulate is necessitated by the relativity principle and Maxwell's equations, as Rovelli and Zeilinger originally envisioned. Accordingly, the fundamental explanans for time dilation, length contraction, quantum superposition, and entanglement is the relativity principle. Phenomenologically speaking, the relativity principle or ``no preferred reference frame'' is ``the equality of all perspectives.'' Thus, quantum entanglement isn't evidence for the violation of intersubjective agreement or the need to abandon factive or objective models of reality, as in relational interpretations of QM. Just the opposite is true. It is evidence that reality is not a fragmented collage of different subjective experiences from different perspectives, it's a comprehensive and coherent integration of those different subjective experiences per the equality of all perspectives.}


\maketitle


\section{Introduction}\label{SectionIntro}

While the quantum foundations community has produced many interpretations of quantum mechanics (QM), none has found consensus support. The problem as articulated by Van Camp \cite{vancamp2011} is that, ``Constructive interpretations are attempted, but they are not unequivocally constructive in any traditional sense.'' He concludes that such constructive interpretations lack explanatory power because they ``must sacrifice some key aspect of the traditional understanding of causal-mechanical explanation.'' In short, despite decades of effort within the quantum foundations community no consensus constructive account of QM has been produced. 

Fuchs \cite{fuchsQMasQI} notes that this impasse is totally analogous to the situation that obtained in the late 1800s regarding length contraction and time dilation before Einstein's 1905 paper on special relativity. In that case, physicists were looking for a constructive account of the observer-independence of the speed of light $c$, including Einstein \cite{einsteinDespair} who wrote:
\begin{quote}
    By and by I despaired of the possibility of discovering the true laws by means of constructive efforts based on known facts. The longer and the more despairingly I tried, the more I came to the conviction that only the discovery of a universal formal principle could lead us to assured results.
\end{quote}

Given this historical precedent, in what Berghofer \cite{BerghoferIQOQI2023} describes as ``Perhaps the paper that can be regarded as the proper beginning of the quantum reconstruction program,'' Rovelli \cite{rovelli1996} suggested we stop trying to \textit{interpret} QM and rather seek to \textit{derive} it in principle fashion:
\begin{quote}
    [Q]uantum mechanics will cease to look puzzling only when we will be able to \textit{derive} the formalism of the theory from a set of simple physical assertions (``postulates'', ``principles'') about the world. Therefore, we should not try to \textit{append} a reasonable interpretation to the quantum mechanics \textit{formalism}, but rather to derive the formalism from a set of experimentally motivated postulates.
\end{quote}
Rovelli was motivated by the success of special relativity where Einstein reconstructed the Lorentz transformations, deriving them from the experimentally motivated light postulate as justified by the relativity principle and Maxwell's equations. Likewise, Zeilinger \cite{zeilingerFoundPrin} writes:
\begin{quote}
    Physics in the 20th century is signified by the invention of the theories of special and general relativity and of quantum theory. Of these, both the special and the general theory of relativity are based on firm foundational principles, while quantum mechanics lacks such a principle to this day. ... I submit that it is because of the very existence of these fundamental principles and their general acceptance in the physics community that, at present, we do not have a significant debate on the interpretation of the theories of relativity. Indeed, the implications of relativity theory for our basic notions of space and time are broadly accepted.
\end{quote}
Hardy \cite{hardy2001} was the first to answer Rovelli's challenge in 2001 with his paper, ``Quantum Theory from Five Reasonable Axioms.'' What Hardy did was to cast quantum mechanics as a probability theory in the manner of classical probability theory. Mackey and Ludwig did likewise in the 1950's using quantum logic \cite{jaeger2018}, but Hardy dropped the quantum logic approach. According to M\"uller (personal correspondence), the ``first fully rigorous, complete reconstructions'' of QM using information-theoretic principles followed in 2011 by Chiribella, D'Ariano \& Perinotti \cite{chiribellaDarianoPerinotti2011} and by Masanes \& M\"uller \cite{masanesMuller2011}. As Berghofer \cite{berghofer2024a} notes, several other reconstructions have been produced, e.g., Daki\'c \& Brukner \cite{dakicBrukner2011}, de la Torre, Masanes, Short, \& M\"{u}ller \cite{torreMasanesShortMuller2012}, Masanes, M\"{u}ller, Augusiak, \& Perez-Garcia \cite{masanesMullerAugPerez2013}, Goyal \cite{goyal2014}, H\"ohn \cite{hohn2017}, and H\"ohn \& Wever \cite{hohnWever2017}. Chiribella \& Spekkens \cite{chiribella1} referred to this as ``the new wave of quantum foundations'' and Grinbaum \cite{grinbaum2006,grinbaum2007} called it a ``paradigmatic shift in the foundations of quantum mechanics.''

What some of these reconstructions and others \cite{jaeger2018,Goyal2026SHPS} showed was that the Hilbert space kinematics of QM with its quantum superposition and entanglement could be derived from Brukner \& Zeilinger's \cite{brukner2009} information-theoretic principle of Information Invariance \& Continuity or its equivalent, e.g., Darrigol's \cite{darrigol2015} ``discreteness'' and ``correspondence'' requirements. In Section \ref{SectionMissing}, we will show that this is in total analogy with how the Lorentz transformation kinematics of special relativity can be derived from the light postulate per Rovelli and Zeilinger's desideratum. 

Despite the success of QRP, philosophers in quantum foundations overall have not been sold on this information-theoretic understanding of QM. Berghofer \cite{berghofer2024b} speculates this lack of interest may be due to the pressure placed on $\psi$-ontic interpretations by QRP. We agree for the reasons Berghofer articulates and we will not expand on that issue here. Instead, we focus on the following two objections leveled against QRP: 
\begin{enumerate}
    \item The information-theoretic principles are ``highly abstract mathematical assumptions without an immediate physical meaning'' \cite{dakicBrukner2011}.
    \item The reconstructions do not contain anything beyond QM, so they do not ``offer more unification of the phenomena than quantum mechanics already does since they are equivalent'' \cite{vancamp2011}. 
\end{enumerate}
In Section \ref{SectionMissing}, we will address Objection 1 by reviewing Darrigol's \cite{darrigol2015} intuitive derivation of the spin-1/2 qubit probabilities from his ``discreteness'' and ``correspondence'' requirements. As Goyal \cite{GoyalPhenomQBism2024} pointed out, there are two steps in QRP: First, reconstruct the quantum formalism and second, interpret the reconstruction. In particular, we will employ ``the four key ideas'' in Goyal's \cite{Goyal2026SHPS} \textit{reconstruction-based interpretative methodology}. 

The first key idea is to deal with the practice of QM as a whole, rather than to focus on its abstract formalism alone. We do this in accord with the second key idea of establishing an \textit{interpretation-free zone} where we ``adopt a \textit{descriptive} – rather than an \textit{explanatory} or \textit{interpretative} – stance.'' In particular, we render the abstract principle of Information Invariance \& Continuity ``intuitively graspable'' \cite{GoyalPhenomQBism2024} using Darrigol's ``discreteness'' and ``correspondence'' requirements in his derivation of spin-1/2 qubit probabilities per Stern-Gerlach (SG) spin measurements. In this context, using empirical and mathematical facts alone, we see that Information Invariance \& Continuity is just qubit superposition and it subsumes the statement that ``everyone measures the same value for Planck's constant $h$, regardless of their spatial orientation relative to the source.'' In fact, Khrennikov's \cite[p. 54]{Khrennikov2025} counterpart to Darrigol's ``discreteness'' requirement is his \textit{quantum action invariance principle} -- ``The quantum of action is the same for all observers, regardless of experimental contexts,'' which he characterizes as ``the epistemic counterpart to Bohr’s quantum postulate (the ontic principle about Nature as it is).'' Since we seek to complete those reconstructions based most fundamentally on Information Invariance \& Continuity (or its equivalent), we have ``distilled what is theoretically relevant'' in our reconstruction-based interpretative methodology, i.e., we have employed Goyal's third key idea. Obviously, we have also used ``\textit{operational analysis} to properly ground all parts of the formalism in experimental practice,'' what Goyal \cite{GoyalFayeBook2026} calls a ``core idea'' for reconstruction-based interpretative methodology. Goyal \cite{Goyal2026SHPS} writes, ``As is well-known, Poincar\'e's penetrating operational analysis was pivotal in Einstein's reconstruction of the Lorentz transformations.'' 

All this will reveal an explanatory lacuna in QRP pointing to the relevance of Objection 2, viz., per these reconstructions, the solution to the mystery of quantum entanglement is to postulate the existence of the equally mysterious qubit superposition. This is like deriving the Lorentz transformations from the light postulate without justifying the observer-independence of $c$ with the relativity principle and Maxwell's equations. As a consequence, one would be solving the mysteries of length contraction and time dilation with the equally mysterious observer-independence of $c$. No one would have accepted Einstein's reconstruction of the Lorentz transformations in that case either. 

In response, per Objection 2, we add a compelling fundamental principle to justify Darrigol's ``discreteness'' requirement and Khrennikov's quantum action invariance principle from outside the practice of QM, i.e., the relativity principle. This completion is incremental and empirically trivial because in Darrigol's derivation of the spin-1/2 qubit probabilities it will be clear that ``discreteness'' is simply the result of the observer-independence of Planck's constant $h$, so it is obviously justified by the relativity principle and Planck's radiation law. That's why H\"ohn \cite{hohn2023} missed the analogy completely when he said, ``Entanglement from complementarity is not as intuitive as the relativity of simultaneity from the relativity principle.'' The proper analogy is that \textit{both} the relativity of simultaneity \textit{and} entanglement follow most fundamentally from the relativity principle. This completion then renders QRP ``a great success'' per Berghofer \cite{berghofer2024a} who wrote, ``if quantum reconstructions can do for quantum mechanics what Einstein did for special relativity, QRP can be considered a great success.'' It also ``offers more unification of the phenomena than quantum mechanics already does'' since it `unifies' the disparate kinematics of Hilbert space and the Lorentz transformations by showing they both follow most fundamentally from the relativity principle \cite{stuckey2025}.

In Section \ref{SectionNM}, we show how our completion of QRP provides ``a reconstruction in terms of experience'' \cite{berghofer2024b} in that the relativity principle or ``no preferred reference frame'' (NPRF) is simply ``the equality of all perspectives'' phenomenologically speaking. In Husserlian phenomenology, objectivity is defined by invariance between perspectives in spacetime physics \cite{Wiltsche2026}. Since phenomenology deals with first-person experience, we're talking about invariance of presentive phenomenal experience from different perspectives in an individual's subjective spacetime model of reality. The relativity principle of physics expands this first-person kinaesthetic variation within a particular subjective spacetime model to third-person anticipatory variation in an objective spacetime model via Poincar\'e transformations between inertial reference frames spanning more than one subjective spacetime model. So completing QRP via NPRF is to ``recognize a relation [between phenomenology and physics] that has been there from the beginning'' \cite{Wiltsche2026}. Consequently, this kinaesthetic and anticipatory invariance for QM means that Bell-inequality-violating correlations at spacelike separation are not evidence for the violation of intersubjective agreement or the need to abandon factive or objective spacetime models of reality, as is sometimes proposed \cite{berghofer2024b,RovelliGreene2024}. Just the opposite is true. The collection of disparate subjective experiences (embodied or inanimately generated) from different perspectives is not to be left fragmented, but integrated according to the equality of all perspectives. Thus, the counterintuitive phenomena of time dilation, length contraction, relativity of simultaneity, quantum superposition, complementarity, and entanglement are kinematic phenomena necessitated by the equality of all perspectives in the integration of disparate subjective experience from different perspectives. This reveals the importance of constructing an objective spacetime model of reality that does not itself require its coordinate origin represent a ``zero point'' of ``lived experience'' \cite{Wiltsche2026}, i.e., the importance of integrating first-person phenomenology with third-person physics to understand reality via ``experiential justification'' \cite{Berghofer2020,einstein1934}.

In doing so, we will introduce ``all at once'' principle explanation per adynamical global contraints in the block universe \cite{StuckeyFoP2008,ourbook,NPRF2024}, as opposed to time-evolved constructive explanation via causal mechanisms in the dynamical universe \cite{ourbook}. This ``situates QM within a metaphysical conception of reality that can be systematically compared with the mechano-geometric-atomistic conception of classical physics,'' i.e., Goyal's \cite{Goyal2026SHPS} fourth key idea in the reconstruction-based interpretative methodology. We then explain why the Pusey–Barrett–Rudolph theorem (PBR) \cite{PBR2012} leads unnecessarily to the belief that Bell-inequality-violating correlations per quantum entanglement entail ``spooky actions at a distance'' (``spukhafte Fernwirkungen'' per Einstein \cite[p. 162]{spooky}). This nonlocality implies ``Relativity can't be right'' \cite{maudlinQMandSR2025}, so we ``must justify the ways of a Deity who is, if not evil, at least extremely mischievous'' \cite[p. 221]{maudlin2011} in giving us a physics that is hideously incoherent. As we will show, just the opposite is true, quantum entanglement obtains due the the relativity principle, which is also fundamental to special relativity, showing that the ``Deity'' is benevolent per strict impartiality giving us a physics that is beautifully coherent. We conclude in Section \ref{SectionConcl}.


\section{QRP: Finding and Fixing its Lacuna}\label{SectionMissing}

Goyal \cite{GoyalPhenomQBism2024,Goyal2026SHPS,GoyalFayeBook2026} points out that the idea of reconstructing a theory of physics traces back to Newtonian mechanics, which was followed by Einstein's reconstruction of the Lorentz transformations via an operational framework. In fact, Einstein \cite{SR1905} specifically defined time using simultaneity in an operational fashion:
\begin{quote}
We have to take into account that all our judgments in which time plays a part are always judgments of \textit{simultaneous events}. [Italics in original.]
\end{quote}
\noindent where his notion of simultaneity was that of the synchronicity of \textit{stationary} clocks, which was established by exchanging light signals \cite{SR1905}:
\begin{quote}
Thus with the help of certain imaginary physical experiments [associated with the exchange of light signals] we have settled what is to be understood by synchronous stationary clocks located at different places, and have evidently obtained a definition of ``simultaneous,'' or ``synchronous,'' and of ``time.'' The ``time'' of an event is that which is given simultaneously with the event by a stationary clock located at the place of the event, this clock being synchronous, and indeed synchronous for all time determinations, with a specified stationary clock. ... \\

\noindent It is essential to have time defined by means of stationary clocks in the stationary system, and the time now defined being appropriate to the stationary system we call it ``the time of the stationary system.'' 
\end{quote}
Essentially, Einstein's operational reconstruction of the Lorentz transformations led to a new interpretation of Maxwell's electrodynamics. 

Goyal \cite{GoyalPhenomQBism2024} and Jaeger \cite{jaeger2018} note that attempts to reconstruct abstract quantum formalism predates the reconstructions of interest here, i.e., those based on information-theoretic principles. Goyal \cite{GoyalPhenomQBism2024} writes:
\begin{quote}
    Recognition of the importance of reconstruction for elucidating the quantum formalism was not lost on the founders. For example, Heisenberg recognized that it would be highly desirable if the quantum formalism could somehow be derived using his uncertainty principle as a key axiom.
\end{quote}
However, the idea that QM deals fundamentally with information goes back at least to Bohr, as summed up by this famous statement by Petersen \cite{merminBohrQuote} about Bohr's belief:
\begin{quote}
    There is no quantum world. There is only an abstract quantum physical description. It is wrong to think that the task of physics is to find out how Nature is. Physics concerns what we can say about Nature.
\end{quote}
Other reconstructions of QM include that of Clifton, Bub \& Halvorson \cite{cliftonBubHalvorson2003} who focused on the algebraic difference between classical and quantum possibility spaces (Boolean versus non-Boolean, respectively) per Heisenberg's commutative versus noncommutative algebra of observables (classical mechanics versus QM, respectively). Bub \& Pitowski \cite{bubpit2010,bubbook,bub2020} compared this difference with the difference between the geometry of Galilean spacetime and Minkowski spacetime \cite{bub2012b}. Rovelli \cite{rovelli1996} also focused on this difference in commutativity by noting that information gained in the measurement of some property of a quantum system is lost when subsequently measuring a noncommutative/non-Boolean complementary property of that system. Herein, we will complete those approaches \'a la Hardy's \cite{hardy2001} that derive the Hilbert space kinematics of QM from Information Invariance \& Continuity (or its equivalent), for reasons that will be clear. 

These quantum reconstructions start with the smallest unit of quantum information, i.e., a quantum ``binary digit'' or ``qubit'' for short, which can be measured in a continuously infinite number of ways always producing two possible outcomes, and build all higher-dimensional Hilbert spaces from that (via tomographic locality \cite{dakicBrukner2011}). Hardy viewed QM as a probability theory and noted that quantum probability theory differs from classical probability theory by just one word in one of his five axioms. Specifically, if you add the word ``continuous'' to one of his axioms, then classical probability theory becomes quantum probability theory.  

To understand this, note that a classical bit can only be measured in one way while a qubit can be measured in a continuous infinity of different ways. For example, a box that may or may not contain a ball is a classical bit and there is only one measurement you can do on it, `open the box' (or its equivalent `weigh the box' or `shake the box', etc.), and there are two possible outcomes, `yes' there is a ball or `no' there is not a ball. Now compare that classical bit with the spin-1/2 qubit per Darrigol \cite{darrigol2015}:
\begin{quote}
There is a continuous infinity of possible measurements of a spin, as there are infinitely many directions of space in which an angular momentum can be measured. In contrast, there are only two possible outcomes for each of these measurements: $+\frac{\hbar}{2}$ and $-\frac{\hbar}{2}$. If the system is found to have the momentum $+\frac{\hbar}{2}$ in a given direction, a subsequent measurement performed in a direction making an angle $\beta$ with the former direction will give either $+\frac{\hbar}{2}$ or $-\frac{\hbar}{2}$.
\end{quote}
As we will see, it is this ``discreteness'' of measurement results with the ``continuity'' of measurement possibilities that leads directly to qubit superposition (whence quantum entanglement). As an empirical fact, ``discreteness'' is part of the interpretation-free zone for this reconstruction. Grinbaum \cite{grinbaum2022} summarized Hardy's approach this way: 
\begin{quote}
Quantum theory is \textit{about} probabilities, a particular composition rule, and a principle of continuity. 
\end{quote}

Since, as Brukner \& Zeilinger \cite{bruknerZeil1999} point out, ``spin-1/2 affords a model of the quantum mechanics of all two-state systems, i.e., qubits,'' we begin by explaining Darrigol's derivation of spin-1/2 qubit probabilities per his requirements of ``discreteness'' (with ``continuity'') and ``correspondence,'' as motivated by Hardy's 2001 paper and inspired by Comte \cite{comte1996}. 

\begin{figure}
\begin{center}
\includegraphics [height = 40mm]{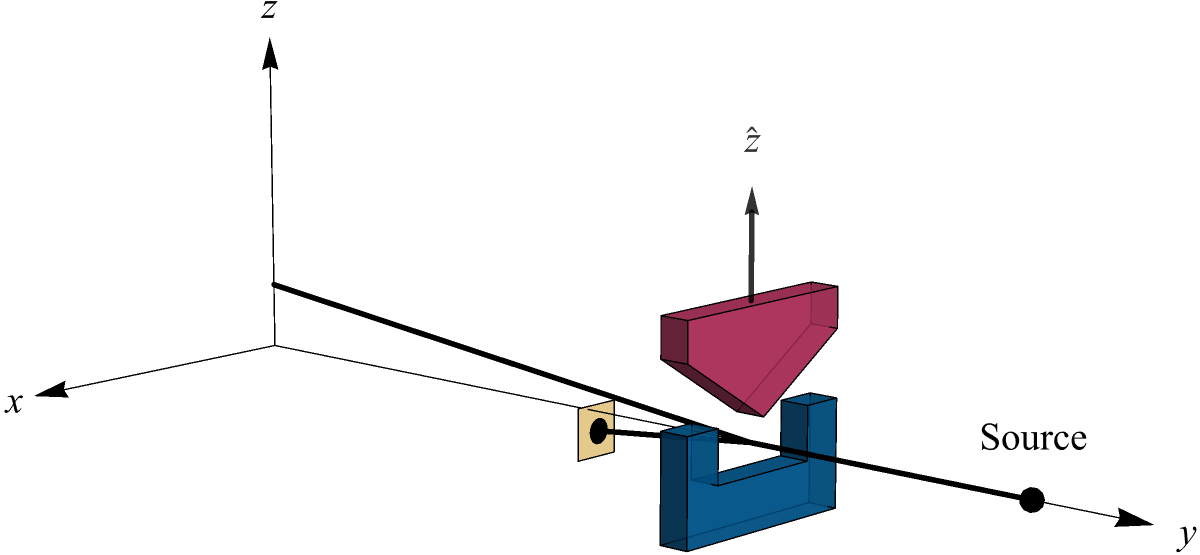}  \caption{In this set up, SG magnets (oriented at $\hat{z}$) and a Source are being used to produce a spin-1/2 qubit in the state $|\psi\rangle = |z+\rangle$ for subsequent measurement.} \label{SGqubit}
\end{center}
\end{figure}

\begin{figure}
\begin{center}
\includegraphics [height = 40mm]{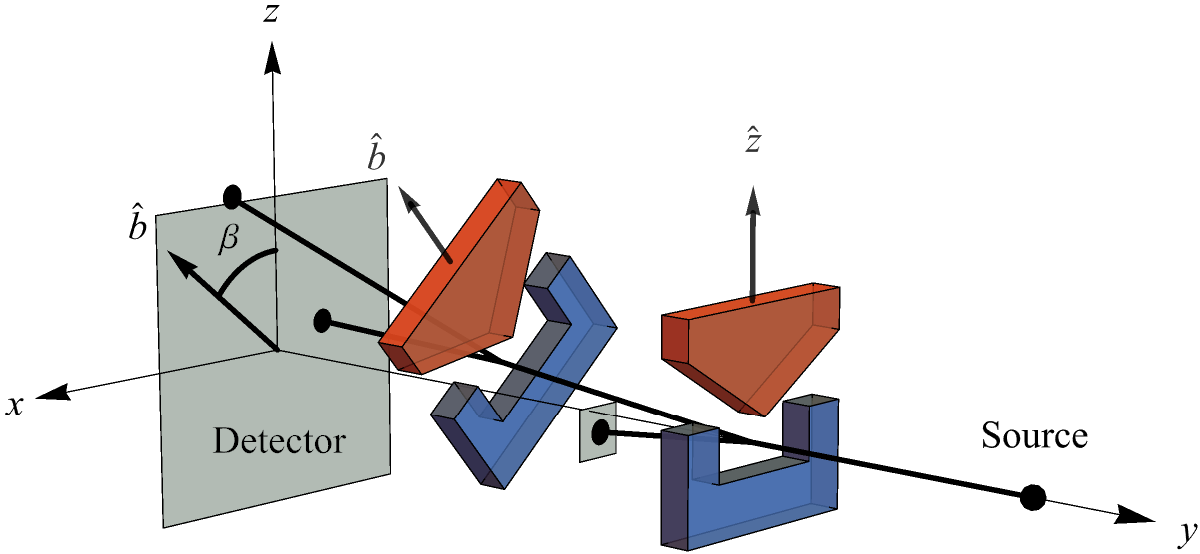}  \caption{In this set up, an SG measurement along $\hat{b}$ is being made on the qubit in Figure \ref{SGqubit}.} \label{SGExp2}
\end{center}
\end{figure}

\begin{figure}[h]
\begin{center}
\includegraphics [width = \textwidth]{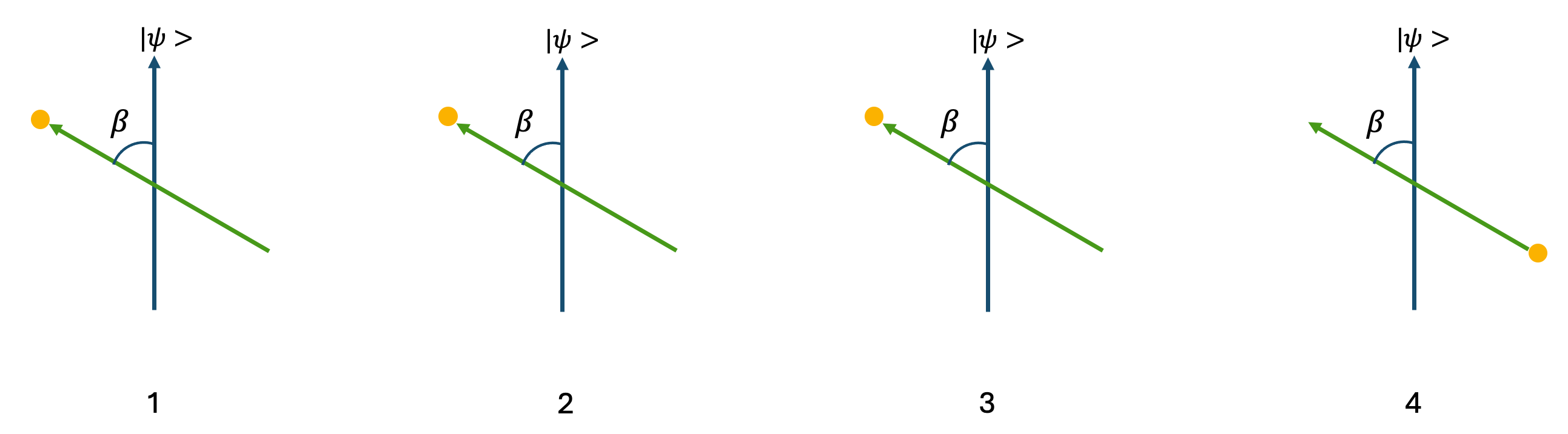} 
\caption{An ensemble of four SG measurement trials for $\beta = 60^{\circ}$ in Figure \ref{SGExp2}. The tilted green arrow depicts the SG measurement orientation $\hat{b}$ and the vertical blue arrow represents the spin angular momentum $+1\hat{z}$ $\left(+\frac{\hbar}{2}\hat{z}\right)$ of our qubit state $|\psi\rangle = |z+\rangle$. The yellow dots represent the measurement outcomes for each trial, `up' (located at arrow tip) or `down' (located at bottom of arrow). The \textit{average} of the $\pm 1$ $\left(\pm \frac{\hbar}{2}\right)$ outcomes equals the projection of the initial spin angular momentum vector $\vec{S} = +1\hat{z}$ in the measurement direction $\hat{b}$, i.e., $\vec{S}\cdot\hat{b} = +1\cos{(60^\circ)}=\frac{1}{2}$.} \label{4DpatternQubit}
\end{center}
\end{figure}

Suppose we use an electron (e.g., valence electron of a silver atom in the Stern-Gerlach (SG) experiment) to create a spin-1/2 qubit in the state $|\psi\rangle = |z+\rangle$ (Figure \ref{SGqubit}) and then we do an SG measurement on it along $\hat{b}$ making an angle of $\beta$ with $\hat{z}$ (Figure \ref{SGExp2}). Since an SG measurement always produces $\pm\frac{\hbar}{2}$ (or $\pm1$ for short) (Darrigol's experimentally motivated ``discreteness'' requirement), repeating the state preparation and $\hat{b}$ spin measurement many times will lead to a distribution of $\pm 1$ outcomes with probabilities $P(+1 \mid \beta)$ and $P(-1 \mid \beta)$ that average to $+1\cos{\left(\beta \right)}$ (Figure \ref{4DpatternQubit}). 

\clearpage

\noindent Per Darrigol \cite{darrigol2015}, this happens: 
\begin{quote}
in order that the average angular momentum in the direction $\beta$ be equal to the projection of the initial angular momentum on this direction. This is so because by a correspondence argument we expect the total angular momentum (or magnetic moment) of a large number of identically prepared, non-interacting spin-particles to behave as the angular momentum of a macroscopic object under measurement (my correspondence requirement $K_2$). 
\end{quote}
Daki\'c \& Brukner \cite{dakic2016} call their version of Darrigol's ``correspondence'' requirement the \textit{closeness requirement} and Comte \cite{comte1996} calls his version the \textit{homogeneity of statistical ensembles}.

\begin{figure}
\begin{center}
\includegraphics [height = 50mm]{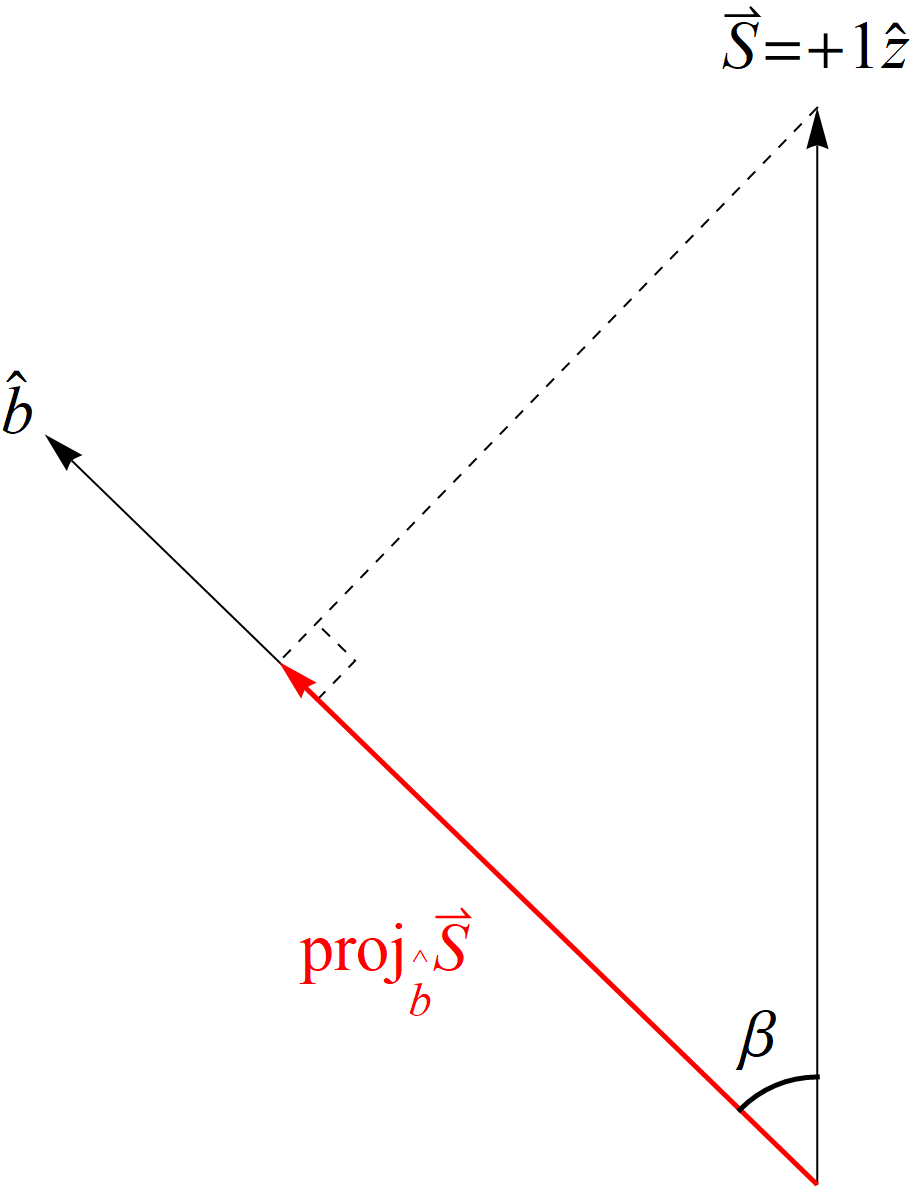} \caption{The spin angular momentum of an electron $\vec{S} = +1\hat{z}$ projected along the measurement direction $\hat{b}$. This does \textit{not} happen with spin angular momentum due to Darrigol's experimentally motivated requirement of ``discreteness.''} \label{Projection}
\end{center}
\end{figure}

Since Darrigol's ``correspondence'' is an empirical fact in accord with simple Newtonian intuition, we consider it part of the interpretation-free zone for this reconstruction. That is, given the spin angular momentum of our spin-1/2 qubit is $+1\hat{z}$ (in units of $\frac{\hbar}{2}$, Figure \ref{SGqubit}), Newtonian mechanics says we should obtain $+1\cos{\left(\beta \right)}$ for the spin angular momentum along $\hat{b}$ by simple vector projection (Figure \ref{Projection}). But, ``discreteness'' says we cannot obtain a fraction of $\pm 1$, so ``correspondence'' says the $\pm 1$ outcomes will \textit{average} to the expected outcome per Newtonian mechanics (Figure \ref{4DpatternQubit}). This gives
\begin{equation}
P(+1 \mid \beta)(+1) + P(-1 \mid \beta)(-1)  = \cos (\beta). \label{AvgProjection}
\end{equation}
That requirement plus normalization, $P(+1 \mid \beta) + P(-1 \mid \beta) = 1$, give 
\begin{equation}
    P(+1 \mid \beta) = \cos^2\left(\frac{\beta}{2}\right) \label{QubitProb+1}
\end{equation}
and
\begin{equation}
    P(-1 \mid \beta) = \sin^2\left(\frac{\beta}{2}\right) \label{QubitProb-1}
\end{equation}
for the spin-1/2 qubit probabilities in accord with QM. 

\begin{figure}
\begin{center}
\includegraphics [height = 60mm]{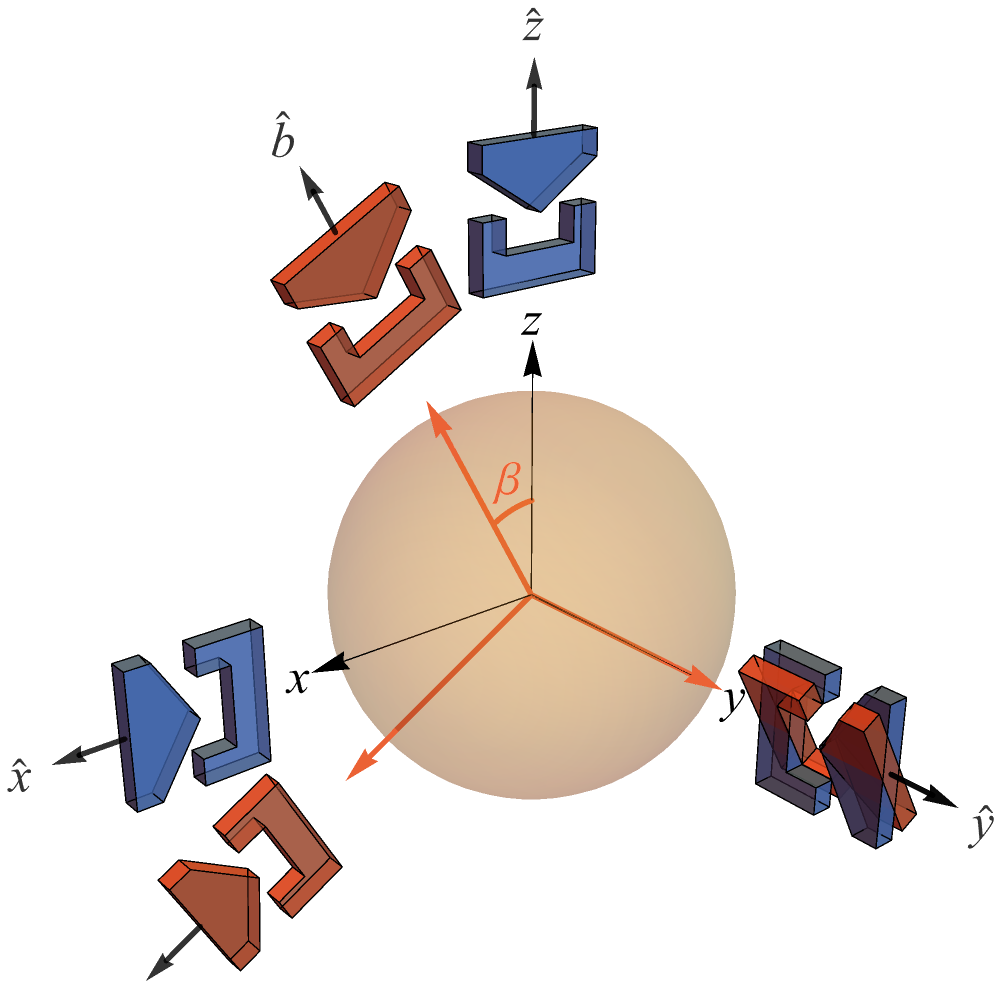}  \caption{Two reference frames established by complementary spin measurements rotated relative to each other (from Brukner \& Zeilinger \cite{bruknerZeil2003}).} \label{SGframes}
\end{center}
\end{figure}

Brukner \& Zeilinger's \cite{brukner2009} information theoretic version of this spin-1/2 qubit superposition is Information Invariance \& Continuity:
\begin{quote}
The total information of one bit is invariant under a continuous change between different complete sets of mutually complementary measurements. 
\end{quote} 
Darrigol's ``intuitively graspable'' result maps to Information Invariance \& Continuity as follows. First, the information contained in our spin-1/2 qubit $|\psi\rangle = |z+\rangle$ is that a $\hat{z}$ spin measurement of $|\psi\rangle$ will produce a $+1$ outcome with 100\% certainty. This amount of information gets `spread out' between $+1$ and $-1$ outcomes as our SG measurement direction $\hat{b}$ continuously changes its angle $\beta$ with respect to $\hat{z}$ such that the distribution $P(+1 \mid \beta)$ and $P(-1 \mid \beta)$ over those $\pm 1$ outcomes always totals 100\%. This corresponds to: ``The total information of one bit is invariant under a continuous change'' in measurement direction $\hat{b}$. The maximum randomization of our initial qubit of information ($+1$ outcome with 100\% probability in the $\hat{z}$ direction) occurs for the complementary measurements, i.e., when $\beta = 90^\circ$, since $P(+1 \mid 90^\circ) = P(-1 \mid 90^\circ) = 50\%$. This is satisfied by $\hat{b} = \hat{x}$ and $\hat{b} = \hat{y}$, so we have a ``complete set of mutually complementary SG measurements'' for $\hat{z}$. Of course, a reference frame for $\hat{b}$ \textit{itself} at any angle $\beta$ can also be constructed in this manner, so we have the second half of Information Invariance \& Continuity: ``between different complete sets of mutually complementary measurements.'' Brukner \& Zeilinger \cite{bruknerZeil2003} depict this with Figure \ref{SGframes}. 

You can see that Darrigol's derivation of the spin-1/2 qubit probabilities begins with ``discreteness,'' but as he points out:
\begin{quote}
    From a physical point of view, [discreteness] is a non-trivial assumption. ... In the present state of this approach, we should probably content ourselves with the insight that quantum discontinuity, if it is admitted as a fundamental feature of the microworld and if it is complemented with natural axioms concerning the relation between micro- and macro-world, necessarily leads to quantum mechanics as we know it.
\end{quote}
In other words, we don't have any justification for ``discreteness,'' but that and ``correspondence'' give us qubit superposition and QM as we know it, so maybe we just have to let the end justify the means? But, to stop the derivation at this point is analogous to deriving the Lorentz transformations from the observer-independence of $c$ without justifying that experimentally motivated fact with the relativity principle and Maxwell's equations. 

As Serway \& Jewett \cite[\S 38.3]{serway}, Norton \cite{nortonSRprinciples}, Knight \cite[p. 1057]{knight}, Young \& Freedman \cite[p. 1218]{youngFreed}, and Rex \& Wolfson \cite[p. 438]{rexWolfson} argue, the relativity principle and Maxwell's equations entail the light postulate of special relativity. That is, if there was only one reference frame for a source in which the speed of light equaled the prediction from Maxwell's equations, then that would certainly constitute a preferred reference frame. Therefore, everyone measures the same value for the speed of light $c$ regardless of their relative motions, as required by the relativity principle and Maxwell's equations. Here is Lorentz's \cite[p. 230]{lorentz} response:
\begin{quote}
    Einstein simply postulates what we have deduced, with some difficulty and not altogether satisfactorily, from the fundamental equations of the electromagnetic field. By doing so, he may certainly take credit for making us see in the negative result of experiments like those of Michelson, Rayleigh and Brace, not a fortuitous compensation of opposing effects, but the manifestation of a general and fundamental principle.
\end{quote}
While Lorentz complains that Einstein ``simply postulates what we have deduced'' (the light postulate) he ends by acknowledging that Einstein made ``us see ... the manifestation of a general and fundamental principle'' (the relativity principle). We think it is readily apparent that the light postulate without the relativity principle is unlikely to have received consensus support. Likewise, we think it is doubtful that QRP will receive consensus support unless its fundamental experimentally motivated ``discreteness'' requirement is justified, so we provide a trivial completion of Darrigol's derivation in total analogy with special relativity.

As Weinberg \cite{weinberg2017} pointed out, the SG measurement of electron spin is a measurement of ``a universal constant of nature, Planck's constant'' $h$, since electron spin is $\pm\frac{\hbar}{2}$. Of course, $h$ is a constant of Nature per Planck's radiation law, just like $c$ is a constant of Nature per Maxwell's equations, so the relativity principle requires the observer-independence of $h$ between the different inertial reference frames related by spatial rotations in Figure \ref{SGframes}. That means we must obtain $\pm 1$ for the SG measurement along $\hat{b}$ for the state $|\psi\rangle = |z+\rangle$ shown in Figure \ref{SGExp2}, instead of $+1\cos{\left(\beta\right)}$ per Newtonian reasoning as shown in Figure \ref{Projection}. Thus, the experimentally motivated requirement of ``discreteness'' (with ``continuity'') whence Information Invariance \& Continuity at the foundation of QRP is due to the observer-independence of $h$, as required by the relativity principle and Planck's radiation law in total analogy with special relativity \cite{NPRF2022,NPRF2024,stuckey2025}. So, Bohr's \cite{Bohr1928} \textit{quantum postulate}, ``which attributes to any atomic process an essential discontinuity, or rather individuality, completely foreign to the classical theories and symbolized by Planck’s quantum of action,'' is equivalent to the observer-independence of $h$. The counterpart to this in Khrennikov's \cite[p. 54]{Khrennikov2025} reconstruction is his \textit{quantum action invariance principle}, ``The quantum of action is the same for all observers, regardless of experimental contexts,'' and he agrees that it is the quantum counterpart to the light postulate. Essentially, Khrennikov replaces Information Invariance \& Continuity (or ``discreteness'' and ``correspondence'' aka qubit superposition) with his principles of quantum action invariance and complementarity. Obviously, the relativity principle completes his reconstruction as well. We conclude this section by showing the relativity principle at work in quantum entanglement. 

\begin{figure}
\begin{center}
\includegraphics [width=\textwidth]{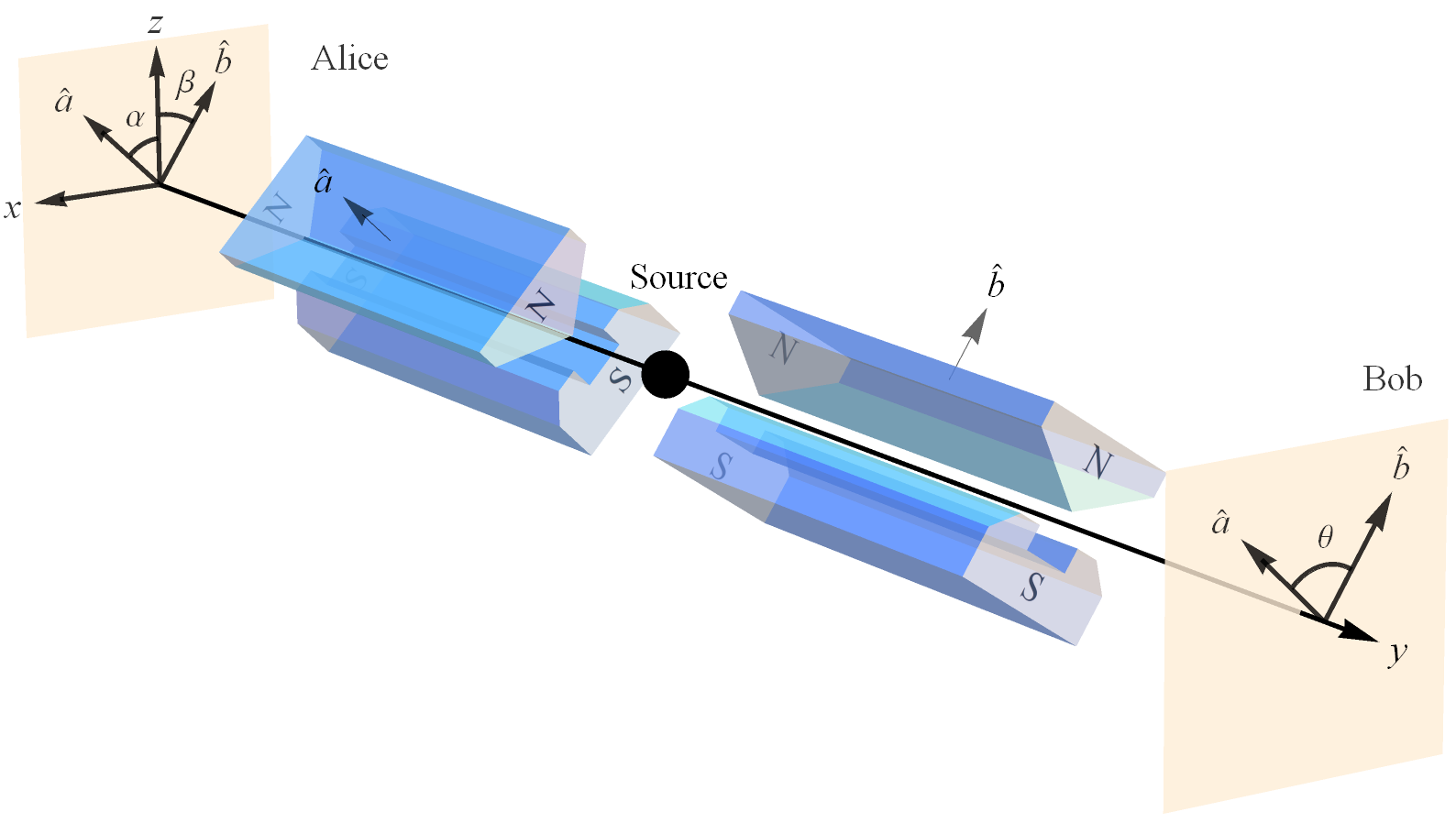} \caption{Alice and Bob making SG spin measurements on a pair of spin-entangled particles.} \label{EPRBmeasure}
\end{center}
\end{figure}

\clearpage

\begin{figure}
\begin{center}
\includegraphics [height = 60mm]{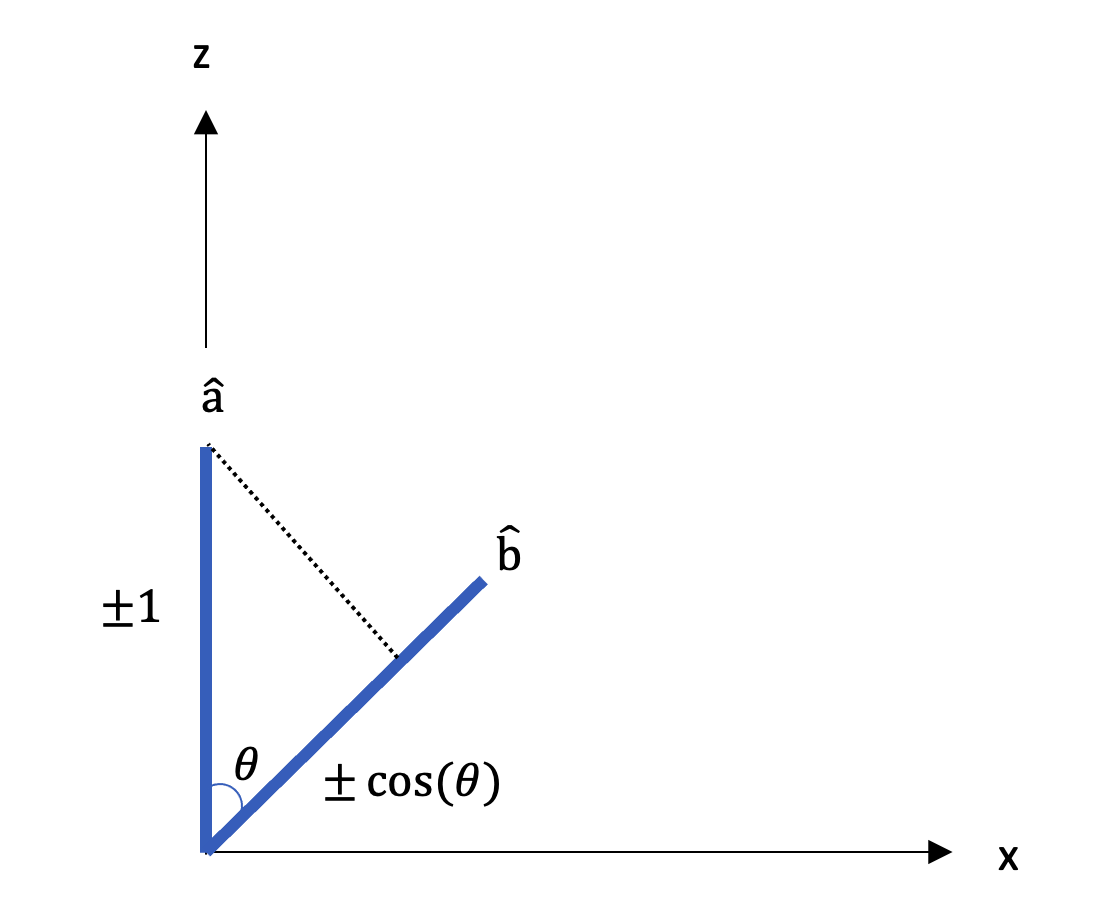} \caption{\textbf{Alice's Perspective}} \label{Alice-View}
\end{center}
\end{figure}

The relativity principle masquerading as the ``discreteness'' requirement is even more obvious in a derivation of the joint probabilities for the Bell spin states. There Darrigol adds ``conservation'' to his ``discreteness'' and ``correspondence'' requirements for Alice and Bob's measurements on a pair of spin-1/2 qubits (Figure \ref{EPRBmeasure}). Since Alice and Bob always obtain the same $\pm 1$ outcomes when their SG magnets are aligned ($\hat{a} = \hat{b}$) per ``conservation'' (in the symmetry plane of a Bell spin triplet state), Alice expects Bob's measurements to produce $\pm \cos{\left(\theta \right)}$ when $\hat{a}$ differs from $\hat{b}$ by angle $\theta$ (Figure \ref{Alice-View}). But, just as in the single particle case, ``discreteness'' means Bob's measurements will always produce $\pm1$ outcomes just like Alice's measurements, so ``correspondence'' tells us Bob's $\pm1$ outcomes will \textit{average} $\pm \cos{\left(\theta \right)}$ as Alice expects from her perspective (her reference frame)
\begin{equation}
2P(+1,+1\mid \theta)(+1) + 2P(+1,-1\mid \theta)(-1) = \cos (\theta) \label{BA+}
\end{equation}
and
\begin{equation}
2P(-1,+1\mid \theta)(+1) + 2P(-1,-1\mid \theta)(-1) = -\cos (\theta). \label{BA-}
\end{equation}

\begin{figure}
\begin{center}
\includegraphics [height = 60mm]{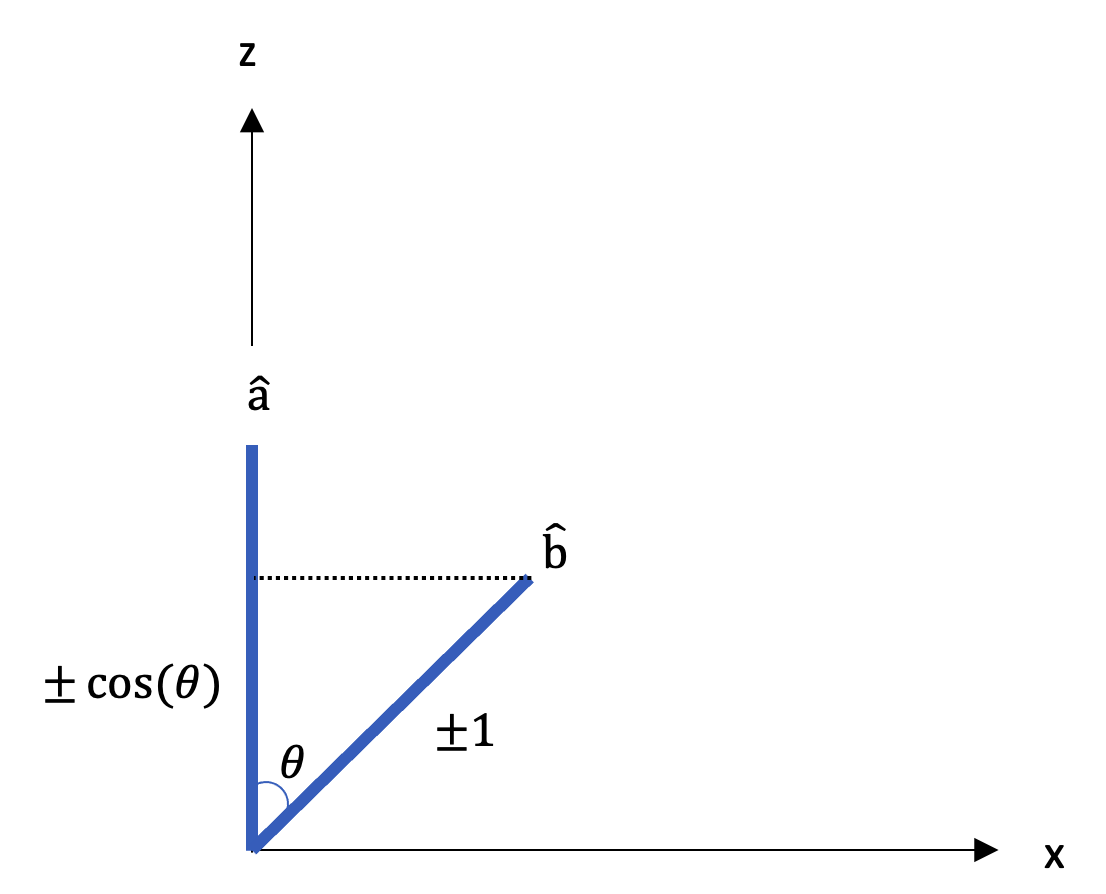} \caption{\textbf{Bob's Perspective}} \label{Bob-View}
\end{center}
\end{figure}

And if there is no preferred perspective (reference frame), then Bob can say exactly the same thing about Alice's results from his perspective (Figure \ref{Bob-View})
\begin{equation}
2P(+1,+1\mid \theta)(+1) + 2P(-1,+1\mid \theta)(-1) = \cos (\theta) \label{AB+}
\end{equation}
and
\begin{equation}
2P(+1,-1\mid \theta)(+1) + 2P(-1,-1\mid \theta)(-1) = -\cos (\theta). \label{AB-}
\end{equation}
These `observer-symmetric', `average-only' conservation equations together with normalization 
\begin{equation*}
P(+1,+1\mid \theta) + P(+1,-1\mid \theta)+P(-1,+1\mid \theta) + P(-1,-1\mid \theta) = 1
\end{equation*}
give the joint probabilities for a Bell spin triplet state in its symmetry plane per QM
\begin{equation}
P(+1,+1 \mid \theta) = P(-1,-1 \mid \theta) = \frac{1}{2} \mbox{cos}^2 \left(\frac{\theta}{2} \right)  \label{QM1jointLike}
\end{equation}
and
\begin{equation}
P(+1,-1 \mid \theta) = P(-1,+1 \mid \theta)  = \frac{1}{2} \mbox{sin}^2 \left(\frac{\theta}{2} \right). \label{QM1jointUnlike}
\end{equation}


\section{Do We Need to Abandon Intersubjective Agreement?}\label{SectionNM}

The completion of QRP via the relativity principle bears prominently on Berghofer's \cite{berghofer2024b} desire for a ``phenomenological reconstruction of quantum theory,'' although with a very different implementation. To appreciate the phenomenological importance of the relativity principle, we start with Einstein's \cite{einstein1936} view that:
\begin{quote}
    ... physics treats directly only of sense experiences and ``understanding'' of their connection. But even the concept of the ``real external world'' of everyday thinking rests exclusively on sense impressions. 
\end{quote}
With this point, Einstein \cite{einstein1934} is simply stating the obvious fact that physics is an empirical science: ``all knowledge about reality begins with experience and terminates in it.'' In that paper \cite{einstein1936}, Einstein also points out that physics begins with the first-person (subjective) experience of interacting ``bodily objects'' like balls, trees, cars, etc., that have worldlines (or worldtubes) in a spacetime diagram (``transtemporal identity'' \cite{GoyalFayeBook2026} or ``identity-through-perspectival-variation'' \cite{Wiltsche2026}). So, each person makes observations personally and with data collection devices that they use to create a first-person (subjective) spacetime model of reality for the interacting bodily objects of their empirical data (of their experience, broadly construed). These subjective spacetime models of reality are anticipatory mental states corresponding to what the perceiver expects to find ``sensuously given'' \cite{Wiltsche2026} via the ``presentive phenomenal character'' \cite{Berghofer2020} of bodily objects should the perceptual origin (``zero point'' \cite{Wiltsche2026}) occupy different perspectives in spacetime. This is the first-person relationship between subject and object per phenomenology that grounds physics. 

The next step in physics (or science more generally) is to assume that everyone else is making similar observations, i.e., having similar experiences (``partly in conjunction with sense impressions which are interpreted as signs for sense experiences for others'' per Einstein). Then we complete ``the game'' of physics by constructing an objective spacetime model of reality that reconciles all the disparate subjective spacetime models. The ``zero point'' of this unificatory objective spacetime model of reality needn't correspond to the perceptual origin of any particular subjective spacetime model, i.e., physics extends phenomenology to the ``third-person'' perspective. This is where relational approaches to QM (e.g., QBism, Healey's pragmatist approach and relational quantum mechanics) run afoul of the game when dealing with ``the greatest mystery in physics'' \cite{aczel2002}, quantum entanglement. 

For example, this is what relational quantum mechanics says about the mystery of entanglement (which Rovelli \cite{RovelliGreene2024} admits ``doesn't take away the strangeness of the phenomenon, the phenomenon is strange and remains strange''). Rovelli acknowledges everything about Alice and Bob's different (presentive phenomenal) experiences from their different perspectives (their individual measurement settings and outcomes in each trial of the experiment), and further he acknowledges that they discover Bell-inequality-violating correlations when meeting later to compare their data. However, since there is no presentive phenomenal experience that subsumes Alice and Bob's first-person experiences of data collection, he simply refuses to play the end game of constructing a unified objective spacetime model of the entire process, i.e., he abandons the necessity of intersubjective agreement on the model because there is no comprehensive objective model, just a fragmented collage of different subjective models. In other words, there is no emphasis on coherently integrating the \textit{totality} of different experiences (both presentive phenomenal and anticipatory) from the different perspectives of the experiencers, because such a model would necessarily include the ``structured anticipation'' \cite{Wiltsche2026} of an otherwise non-existent presentive phenomenal perspective on their disparate past subjective experiences. Bitbol \cite{BitbolCloserTruth2026} agrees saying the Bell-inequality-violating correlations don't exist until Alice and Bob's results have been shared, i.e., ``There is no correlation until there is a measured correlation.'' QBism, for example, emphasizes the prediction of our individual future subjective experiences, so perhaps the QBist doesn't care so much about such a complete objective model. But, don't we want comprehensively coherent explanations of our past experiences as well? And without such explanations, aren't we losing a strong predictor of future experiences via induction? If this omission of explanation isn't bad enough, relational views of QM can and do lead to different observers legitimately experiencing (in a presentively phenomenal sense) different outcomes for one and the same measurement in the same world! Essentially, ``we might warn that an interpretation of a scientific theory should not cut off the branch on which science is sitting'' \cite{BergGoyalWilt2021}.

Thankfully, QRP does not force us into the undesirable omission or violation of intersubjective agreement on the model. In fact, our completion of QRP reveals the key to explaining our past and predicting our future subjective experiences is ``the equality of the perspectives of the experiencers'' when constructing an intersubjectively-agreed-upon and comprehensive, objective spacetime model of reality. In other words, reality is not a fragmented collage of different subjective experiences (both presentive phenomenal and anticipatory) from different perspectives, it's a comprehensive and coherent integration of those different subjective experiences per the equality of all perspectives relative to all perceptual origins. We believe this marks ``progress in ameliorating the seemingly stark opposition between the subject-relative life-world and the objective world of mathematized science,'' and it clarifies the ``constitutional status of the mathematical machinery of quantum mechanics'' per Islami \& Wiltsche \cite{Wiltsche2025}. This vindicates their belief that QRP might ``serve as a mediator between the lofty heights of the abstract quantum formalism and lower-level principles that are more amenable to ... phenomenological analysis.'' And to accomplish this, we need only give up the constructive bias that forced the quantum interpretation program into its stagnation, and fully commit to the principle approach to QM that launched QRP per Rovelli and Zeilinger originally (Section \ref{SectionIntro}).


\subsection{Constructive versus principle approaches to QM} \label{SubsectionMaudlin}

Concerning constructive theories, Allori \citep{allori2023} writes:
\begin{quote}
    Constructive theories allow one to understand the phenomena compositionally and dynamically: macroscopic objects are composed of microscopic particles, and the macroscopic behavior is completely specified in terms of the microscopic dynamics. Therefore, the type of explanation these theories provide is bottom-up, rather than top-down. 
\end{quote}
If we insist on rendering QM a constructive theory, then we are led to Maudlin's conclusion that, ``Basically, what Bell proved is that Relativity can't be right'' \cite{maudlinQMandSR2025}, so we ``must justify the ways of a Deity who is, if not evil, at least extremely mischievous'' \cite[p. 221]{maudlin2011}. In that case, the violation of Bell's inequality due to quantum entanglement is evidence that:
\begin{itemize}
    \item Special relativity is wrong and in conflict with QM because 
    \item There is a preferred perspective on reality due to 
    \item A ``Deity'' that is evil or extremely mischievous giving us
    \item A physics that is hideously incoherent.
\end{itemize}

\begin{figure}
\begin{center}
\includegraphics [height = 80mm]{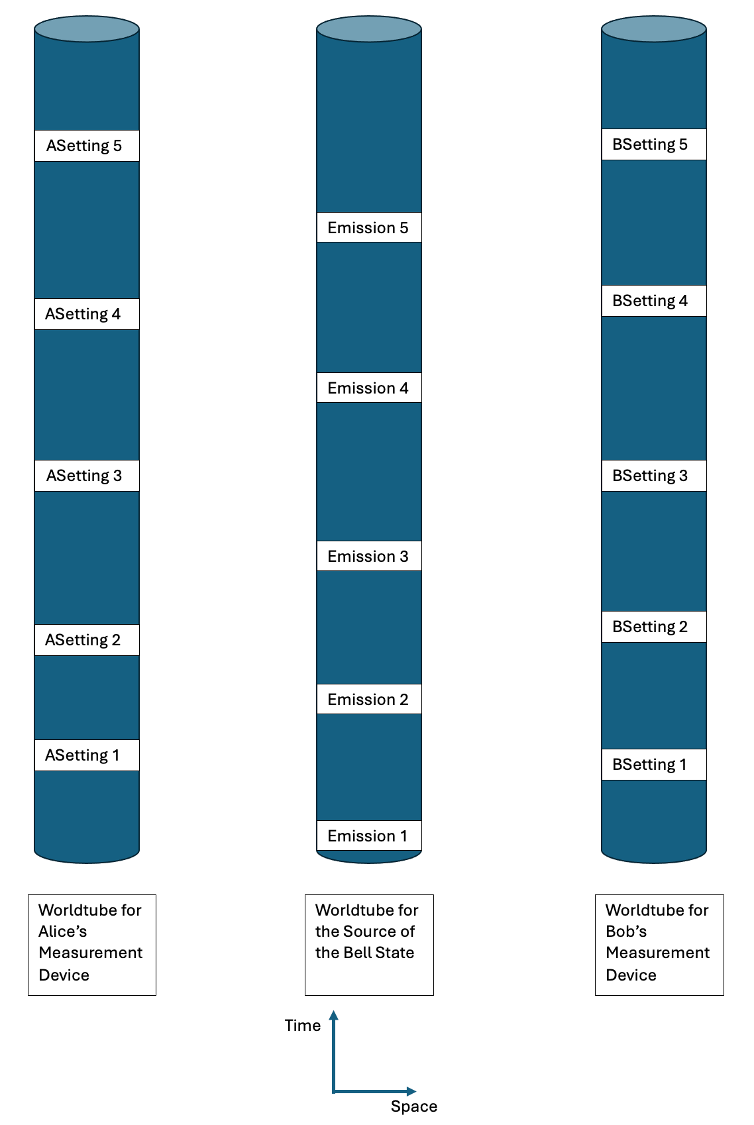}  \caption{Alice and Bob choose their measurement settings randomly and independently for each trial/emission event of the experiment and the Bell state Source is faithfully reproduced in each trial. This spacetime diagram is in accord with the adynamical global constraint NPRF + $c$.} \label{AGC1}
\end{center}
\end{figure}

\begin{figure}
\begin{center}
\includegraphics [height = 80mm]{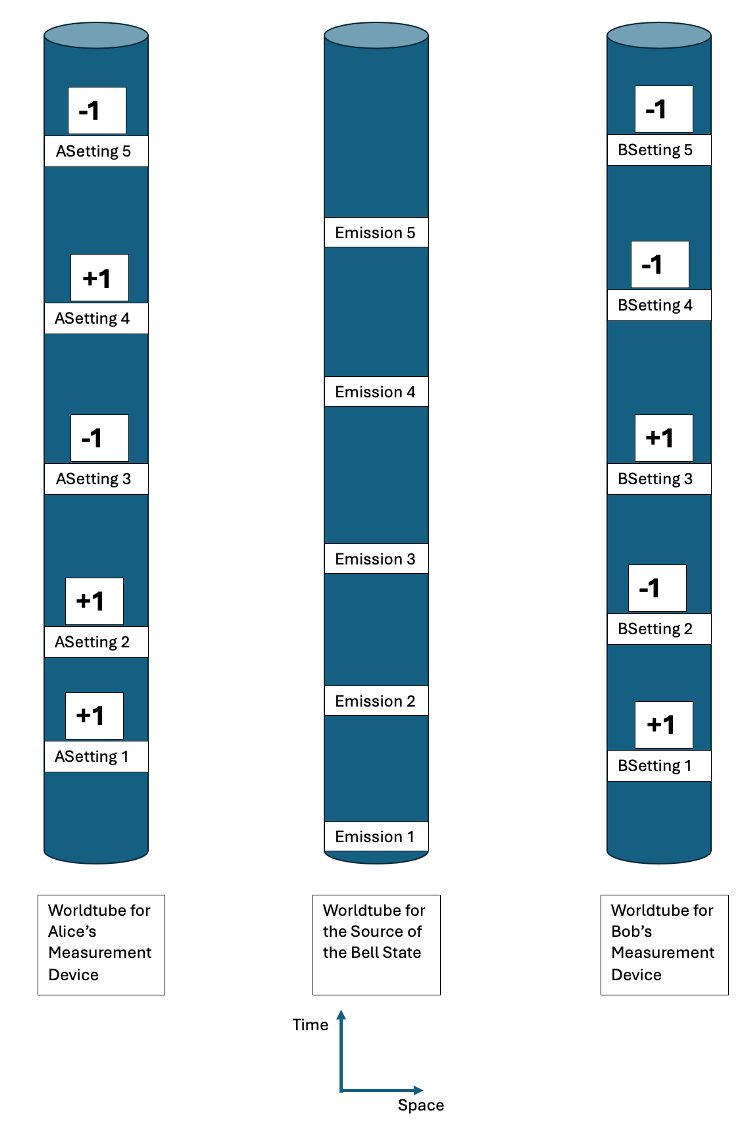}  \caption{The joint probabilities per our acausal global constraint NPRF + $h$ provide an ``all at once'' distribution of outcomes throughout spacetime for Alice and Bob's measurement choices over the Bell state Source in question.} \label{AGC2}
\end{center}
\end{figure}

Alternatively, we can accept QM as a ``greatly successful'' principle theory per QRP as grounded in the relativity principle. Concerning principle theories, Allori \citep{allori2023} writes: 
\begin{quote}
    Principle theories, also called kinematic theories, are formulated in terms of principles, which are used as constraints on physically possible processes: they exclude certain processes from physically happening. In this sense, principle theories are top-down: they explain the phenomena identifying constraints the phenomena need to obey. They are `kinematic' theories because the explanations they provide do not involve dynamical equations of motion and they do not depend on the interactions the system enters into. 
\end{quote}
In this case, there is no wavefunction realism (the wavefunction is not a thing in itself), the wavefunction isn't representing one's knowledge about a physical system as determined by some hidden variables, and it doesn't merely ``represent one’s degrees of belief about, or objective degrees of epistemic justification regarding, the contents of one's future experiences'' \cite{berghofer2024b}. Rather, the wavefunction represents the spatiotemporal distribution of quantum events in the context of the worldtubes of bodily objects in the objective spacetime model of reality, as constrained by the perspectival equality of all experiencers (broadly construed). That is, the relativity principle demands the observer-independence of $c$ whence Minkowski space kinematics constraining the configuration of worldtubes of bodily objects in spacetime (Figure \ref{AGC1}) and the relativity principle demands the observer-independence of $h$ whence Hilbert space kinematics constraining the distribution of quanta among those worldtubes (Figure \ref{AGC2}).

This is a version of ``all at once'' explanation \cite{StuckeyFoP2008,ourbook,NPRF2024}, as inspired by this Geroch \cite[p. 20--21]{geroch} quote:
\begin{quote}
    There is no dynamics within space-time itself: nothing ever moves therein; nothing happens; nothing changes. In particular, one does not think of particles as moving through space-time, or as following along their world-lines. Rather, particles are just in space-time, once and for all, and the world-line represents, \textit{all at once}, the complete life history of the particle. [Italics added.]
\end{quote}
Constructive versions of ``all at once'' explanation have been used by Liu and Wharton for retrocausality \cite{wharton3,WhartonLiu2022,adlam2022Synthese}, Esfeld and Gisin for Bell flash ontology \cite{EsfeldGisin2013}, Hance, Hossenfelder and Palmer for superdeterminism \cite{supermeasured2022}, and Adlam and Rovelli for relational quantum mechanics \cite{AdlamRovelli}. Of course, the `big' measurement problem \cite{pitowski2007,bubpit2010} is deflated trivially by this ``all at once'' explanation, since the `big' measurement problem is to explain ``how individual measurement outcomes come about dynamically.'' Obviously, that question is a nonstarter when the formalism is used to produce distributions of outcomes atemporally in a spatiotemporally global fashion (Figure \ref{AGC2}).   

Accordingly, quantum entanglement is not a dynamical effect due to some nonlocal or superdeterministic or retro- causal mechanism. It is a kinematic fact that follows necessarily from the observer-independence of $h$, as required by the relativity principle and Planck's radiation law (NPRF + $h$). This is totally analogous to special relativity where length contraction and time dilation are not dynamical effects due to some causal mechanism like the luminiferous aether. They are kinematic facts that follow necessarily from the observer-independence of $c$, as required by the relativity principle and Maxwell's equations (NPRF + $c$). In that case, the violation of Bell's inequality due to quantum entanglement is evidence that:
\begin{itemize}
    \item There is no preferred perspective on reality so 
    \item The relativity principle grounds both special relativity and QM because 
    \item The ``Deity'' is benevolent per strict impartiality giving us
    \item A physics that is beautifully coherent.
\end{itemize}
We know where we're placing our bets.


\subsection{Is QRP in tension with the PBR theorem?}

Of course, since QRP is just a reconstruction of the formalism of QM, it cannot, in and of itself, be in tension with any theorem that assumes the formalism of QM is exactly right, as the Pusey–Barrett–Rudolph theorem (PBR) \cite{PBR2012} does. So, Berghofer's \cite{berghofer2024a} worry that QRP is in tension with PBR is motivated by some additional assumptions that are not germane to QRP. In what follows, we explain what PBR means in terms of QRP in order to expose the unnecessary constructive bias leading to ``a physics that is hideously incoherent'' and this worry. 

PBR assumes there are some hidden variables (called the \textit{ontic state}) associated with each pure state in Hilbert space. What PBR proved was that each pure state has its own hidden variables, i.e., it doesn't share any hidden variables with any other pure state. Since the formalism of QM does not assume the existence of hidden variables, this is an assumption beyond QRP. Let's look at QRP as completed here and see what PBR is actually telling us.

\begin{figure}
\begin{center}
\includegraphics [height = 40mm]{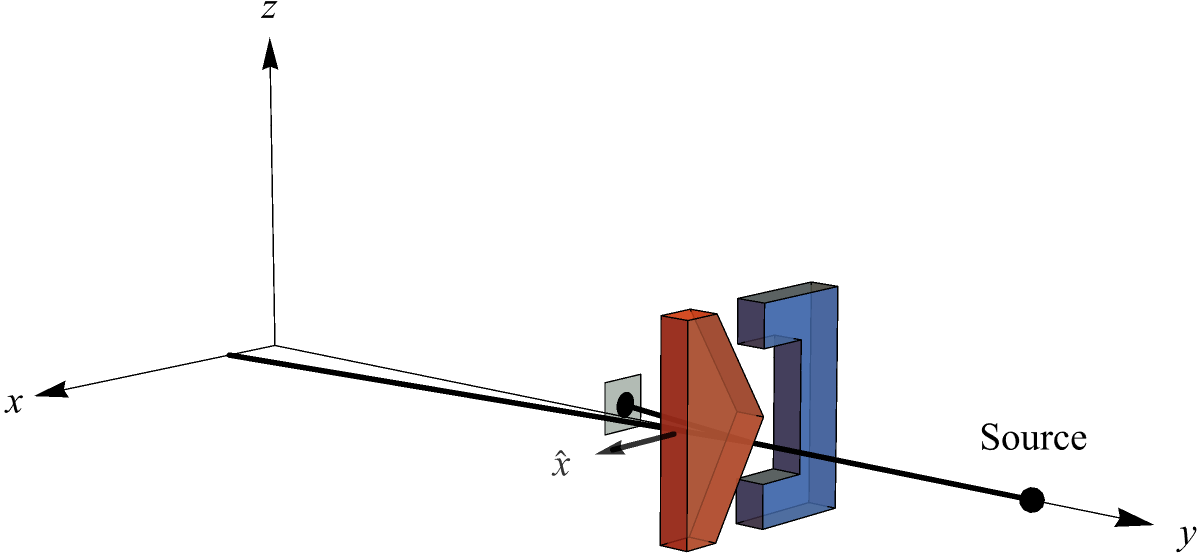}  \caption{In this set up, SG magnets (oriented at $\hat{x}$) and a Source are being used to produce a spin-1/2 qubit in the state $|\psi\rangle = |x+\rangle$ for subsequent measurement.} \label{SGqubitX}
\end{center}
\end{figure}

Consider two spin-1/2 qubits in the pure states $|\psi\rangle = |z+\rangle$ (Figure \ref{SGqubit}) and $|\psi\rangle = |x+\rangle$ (Figure \ref{SGqubitX}). As these qubits are measured along various $\hat{b}$ (Figure \ref{SGExp2}) between $\hat{z}$ and $\hat{x}$ the distributions will not be equal unless $\hat{b}$ happens to lie exactly midway between $\hat{z}$ and $\hat{x}$, since the distributions are determined by the projection angles for $\hat{b} \cdot \hat{z}$ and $\hat{b} \cdot \hat{x}$, respectively. Accordingly, it is not surprising that PBR concludes, using the formalism of QM, that there can be no overlap between the hypothetical ontic states for these two pure states. Thus, there is no need to posit the existence of hidden variables in constructive fashion per the incompleteness of QM, since it is easy to see why the spatiotemporal distributions (Eqs. \ref{QubitProb+1} and \ref{QubitProb-1}) obtain in a globally contextual fashion per the relativity principle. As shown in Subsection \ref{SubsectionMaudlin}, in order to interpret QM constructively one is positing its incompleteness leading to nonlocality that proves special relativity contradicts QM and ``can't be right'' \cite{maudlinQMandSR2025}, i.e., ``a physics that is hideously incoherent.'' For example, here is how Leifer \cite{leifer2012} links PBR and quantum nonlocality via the Bell spin triplet state of Section \ref{SectionMissing}.  

When Alice chooses an SG measurement $\hat{a}$ for her qubit (what Leifer calls the ``first qubit'' measurement), the two possible outcomes are represented by a pair of orthogonal vectors in Hilbert space, call them $|a+\rangle$ and $|a-\rangle$. Since her qubit and Bob's (Leifer's ``second qubit'') are entangled in a Bell spin triplet state, she knows that whatever she gets as her outcome is what Bob will get for his outcome for that same measurement, i.e., if $\hat{b} = \hat{a}$ in Figure \ref{EPRBmeasure}, per the ``conservation'' requirement. Thus, as explained in Section \ref{SectionMissing}, from Alice's perspective Bob's state will be either $|\psi_B\rangle = |a+\rangle$ or $|\psi_B\rangle = |a-\rangle$ (Figure \ref{Alice-View}) for determining his distribution of outcomes in the basis $|b+\rangle$ and $|b-\rangle$ of his SG measurement $\hat{b}$ per `average-only' conservation (Eqs. \ref{BA+} and \ref{BA-}) following from the ``correspondence'' requirement. In this fashion, Alice's choice of the ``first qubit'' measurement is ``steering'' the subsequent state of Bob's ``second qubit.'' Since this is true for any direction $\hat{a}$ and all these pure states have distinct ontic states, the ontic state of Bob's ``second qubit'' is determined by Alice's choice of $\hat{a}$. Finally, her choice of $\hat{a}$ and Bob's SG measurement outcome can be spacelike related, so we have quantum nonlocality.

Notice that the hypothetical hidden variables of the ontic state play an important role in the causal explanation of how the ``conservation'' requirement is dynamically `enforced' in this constructive account of the joint probabilities for the Bell spin triplet state (Eqs. \ref{QM1jointLike} and \ref{QM1jointUnlike}). Notice also that there is nothing in QRP necessitating any of this constructive explanation leading to ``a physics that is hideously incoherent.'' Just the opposite is true. All of this ``spooky actions at a distance'' is the result of trying to use the principle theory of QM to produce a constructive explanation of the Bell-inequality-violating correlations of quantum entanglement. Here instead is the principle explanation per our completion of QRP ``giving us a physics that is beautifully coherent.''

\begin{figure}
\begin{center}
\includegraphics [height = 40mm]{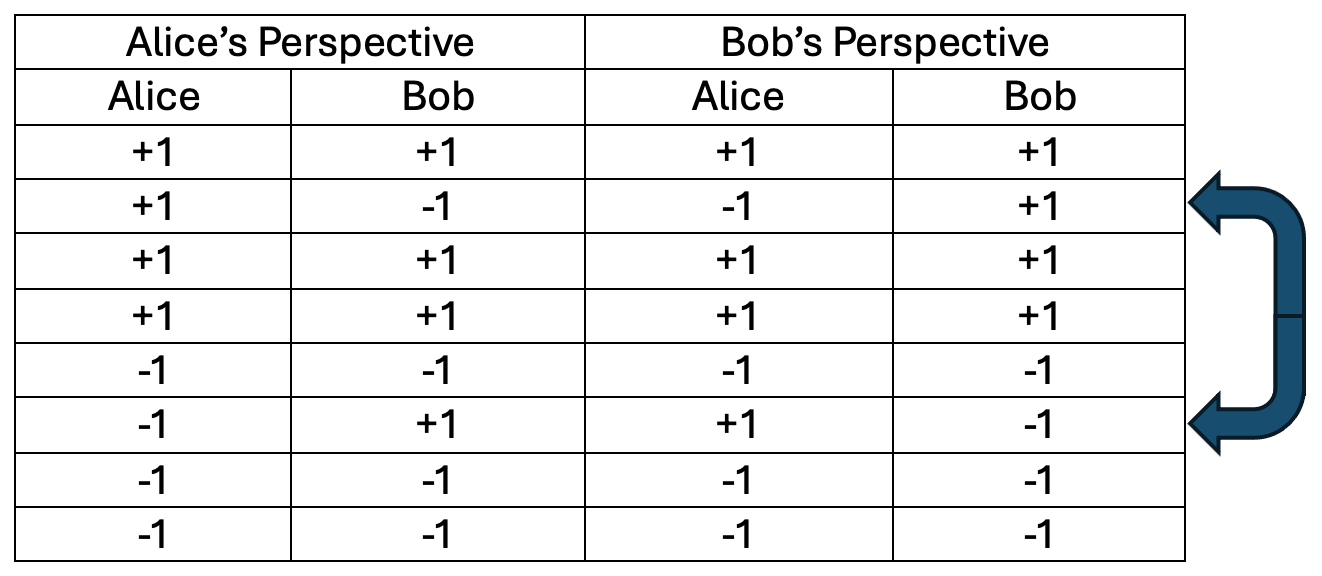}  \caption{Example collection of eight data pairs when Alice and Bob's measurement settings in the symmetry plane for a Bell spin triplet state differ by $60^{\circ}$. Alice partitions the data according to her $\pm 1$ results to show that Bob's measurement outcomes only \textit{average} the required $\pm\frac{1}{2}$ for conservation of spin angular momentum from her perspective. But, when Bob partitions the data according to his $\pm 1$ results (by switching rows 2 and 6 shown with dark blue arrows) he can show it is \textit{Alice's} measurement outcomes that only average the required $\pm\frac{1}{2}$ for conservation of spin angular momentum from his perspective.} \label{AliceBobData}
\end{center}
\end{figure}

According to the equality of all perspectives, Bob can make the same claim about ``steering'' the state of Alice's qubit per his choice of SG measurement $\hat{b}$ so that $|\psi_A\rangle = |b+\rangle$ or $|\psi_A\rangle = |b-\rangle$ (Figure \ref{Bob-View}) for determining her distribution of outcomes in the basis $|a+\rangle$ and $|a-\rangle$ of her SG measurement $\hat{a}$ per `average-only' conservation from his perspective (Eqs. \ref{AB+} and \ref{AB-}) following from the ``correspondence'' requirement. So, who is ``steering'' whose qubit state? No one. Quantum entanglement is not a dynamical effect due to some nonlocal causal mechanism. It is a kinematic fact due to the observer-independence of $h$ that follows necessarily from the relativity principle and Planck's radiation law. Bob is no more ``steering'' Alice's qubit state than Alice is ``steering'' Bob's in this principle explanation (Figure \ref{AliceBobData}). Consequently, special relativity is not in ``tension'' with QM and it is not wrong. Again, just the opposite is true. Special relativity and QM are both principle theories providing principle explanations for quantum superposition, entanglement, complementarity, length contraction, time dilation, and the relativity of simultaneity with the exact same fundamental explanans, i.e., the relativity principle. Consequently, Bell-inequality-violating correlations per quantum entanglement don't impugn special relativity, they corroborate it by providing another important consequence of the equality of all perspectives. 

Thus, the results of the PBR calculations done with the QM formalism have nothing to do with hidden variables. The PBR results are simply showing how the pure states of QM are referring to unique spatiotemporally contextually defined configurations of interacting bodily objects per the exchange of quanta. For example, the spin-1/2 qubit in the pure state $|\psi\rangle = |z+\rangle$ (Figure \ref{SGqubit}) is obviously distinct from the spin-1/2 qubit in the pure state $|\psi\rangle = |x+\rangle$ (Figure \ref{SGqubitX}) and this distinction per their different spatial orientations relative to the Source leads precisely and intuitively to their different distributions for an SG measurement $\hat{b}$ in principle fashion (giving a ``beautifully coherent physics''). Contrast that with having to posit different hypothetical hidden ``ontic states'' for the intrinsic properties of electrons that would still have to be coupled with some (unknown) causal mechanism to produce the QM distributions (giving a ``hideously incoherent physics''). Care to place your bet?

\section{Conclusion}\label{SectionConcl}

Herein, we completed those reconstructions in QRP that are based most fundamentally on ``discreteness'' (with ``continuity'') and ``correspondence,'' i.e., Information Invariance \& Continuity or qubit superposition. We did this by showing how the experimentally motivated ``discreteness'' requirement at the foundation of these reconstructions of the Hilbert space kinematics of QM is the result of the observer-independence of $h$, which makes their completion with the relativity principle incremental and trivial. Thereby, QRP has done ``for quantum mechanics what Einstein did for special relativity,'' so it ``can be considered a great success,'' and it exactly satisfies Rovelli and Zeilinger's desideratum for understanding QM in principle fashion \'a la special relativity. 

By focusing on the practice of QM as a whole and operationally grounding its formalism in experimental practice, it is easy to see that these reconstructions are based on mathematical and empirical facts, so they reside entirely in the interpretation-free zone of the reconstruction-based interpretative methodology. Only our completion with the relativity principle, which we promote to nomological rather than mere empirical status, can be said to reside outside the interpretation-free zone. Consequently, ``no preferred reference frame'' (NPRF) provides the fundamental explanans for quantum superposition, complementarity and entanglement. Of course, this is exactly the status that the relativity principle has in special relativity, so we see that the counterintuitive phenomena of time dilation, length contraction, the relativity of simultaneity, quantum superposition, complementarity, and entanglement are all ultimately just consequences of ``the equality of all perspectives.'' 

Thus, phenomenologically speaking, our completion of QRP reveals a reality that is not a fragmented collage of different subjective experiences from different perspectives, it's a comprehensive and coherent integration of those different subjective experiences per the equality of all perspectives. Accordingly, we see that QRP can ``serve as a mediator'' between abstract QM formalism and the life-world of everyday experience per phenomenological analysis. 

Despite its ``great success'' in answering Wheeler's \cite{wheeler1986,wheeler2004} ``Really Big Question,'' ``Why the quantum?'', this completion deals purely with the kinematics of QM, so it does not address the `big' measurement problem from the dynamics of QM. As it turns out, our principle constraint-based explanation for QM allows it to be ``systematically compared with the mechano-geometric-atomistic conception of classical physics.'' According to this ``all at once'' explanation, NPRF + $c$ and NPRF + $h$ are adynamical global constraints on the configuration of worldtubes for bodily objects in spacetime and the distribution of quanta among those worldtubes, respectively. As a consequence, the `big' measurement problem was trivially deflated. Finally, we used this spatiotemporally contextual principle explanation to deflate the worry that QRP is in tension with the PBR theorem.

\bibliography{biblio} 
\end{document}